# Bond relaxation, electronic and magnetic behavior of 2D metals structures Y on Li(110) surface

Maolin Bo,[a] Li Lei,[a] Chuang Yao,[a] Zhongkai Huang,[a] Cheng Peng,[a]* Chang Q. Sun[a,b]*,

[a]*Key Laboratory of Extraordinary Bond Engineering and Advanced Materials Technology (EBEAM) of Chongqing, Yangtze Normal University, Chongqing 408100, China*

[b]*NOVITAS, School of Electrical and Electronic Engineering, Nanyang Technological University, Singapore 639798, Singapore*

*E-mail: 20090008@yznu.cn; ecqsun@ntu.edu.sg*

## Abstract

We investigated the bond, electronic and magnetic behavior of adsorption Yttrium(Y) atoms on Lithium (Li)(110) surface using a combination of Bond-order-length-strength(BOLS) correlation and density-functional theory(DFT). We found that adsorption Y atoms on Li(110) surfaces form two-dimensional (2D) geometric structures of hexagon, nonagon, solid hexagonal, quadrangle and triangle. The consistent with the magnetic moment are 6.66 μB, 5.54 μB, 0.28 μB, 1.04 μB, 2.81 μB, respectively. In addition, this work could pave the way for design new 2D metals electronic and magnetic properties.

1. Introduction

The ultrathin metallic materials have received extensive research because of their attractive, and applications in superconductor[1], catalysis[2, 3], biosensing[4, 5], magnetic property[6, 7] and so on[8]. Although there have been a few reviews focusing on ultrathin metal materials[9], however an atomic layer of metallic electronic structure and magnetic properties are still lacking[10, 11]. Because of the skin of the nonbonding electrons polarization is makes an atomic layer metals unstable, and hence difficult to achieve. Thus, there is only a few reports have demonstrated an atomic layer thin of 2D metal structure formation over substrates.

The Li and Y metal are both the less electronegative. Due to Li metal is the proximity of its valence electron close to nucleus, which the remaining two electrons are in the 1$s$ orbital, much lower in energy, and do not participate in chemical bonds[12]. Thus, the chemical bond between the Y and Li metal is weak without chemical reaction taking place when interacted. When the Y metal and Li substrates chemical bond energy and length are like van der Waals force of layer and layer interacted, and hence the 2D structure of Y metal formation over Li metal substrates.

Therefore, we have performed BOLS notation[13] and DFT calculations of adsorption Y atoms on Li(110) surfaces. We mainly focus on magnetic properties, electronic structure and bond relaxation of an atomic layer Y metal on Li(110) surface. This paper aims to provide a picture of the BOLS correlation research on 2D metal materials and share some perspectives of 2D metal structure formation over substrates.

2. Principles

**2.1 DFT calculation methods**

We calculate the electronic and magnetic distribution of an atomic layer Y metal on Li(110) surface using first principles. The optimal geometric configurations are shown in the **Fig. 1**. The Vienna *ab initio* simulation package and the plane-wave pseudopotential are used in the calculations. We also employed the Perdew–Burke–Ernzerhof exchange-correlation potentials with spin-orbit coupling.[14] The plane-wave cut-off is 400 eV. The Brillouin zone is calculated with special *k*-points generated in an $8 \times 8 \times 1$ mesh grid. The vacuum space of all surfaces is 18 Å. In the calculations, the energy converged is $10^{-5}$ eV and the force on each atom converged to be <0.01 eV/Å.

**2.2 Bond-order-length-strength (BOLS) notation**

According to the BOLS correlation, XPS experiment is using the determinations between surface and bulk atoms by energies shifts difference. It was realized that the energy shift depend on the atomic coordination number (CN) *z*. For a given DFT calculation of the energy shifts, we have the relation of energy level *v* of the energy shifts $\Delta E_v(i)$ and the atomic CN *z* as follows[15]:

$$\begin{cases} z = \dfrac{12}{\left\{8\ln\left(\dfrac{2\Delta E_v'(i) + \Delta E_v(12)}{\Delta E_v(12)}\right) + 1\right\}} \quad \text{(DFT)} \\ E_v(i) = \Delta E_v'(i) + E_v(12) \end{cases}$$

(1)

The $\Delta E_v'(i) = E_v(i) - E_v(12)$ is represents the energy shift of an atom in an ideal bulk. We achieve the quantitative information of bulk shifts $\Delta E_v(12) = 3.661$ eV which are from the Ru[16] and Sc[17] surface for reference Y metals with the 2 eV Gaussian width and the same HCP structure. Then We have

$$\begin{cases} z = 12 \Big/ \left\{8 \ln\left(\dfrac{2\Delta E_v'(i) + 3.661}{3.661}\right) + 1\right\} \\ E_v(i) = \Delta E_v'(i) + E(12) \end{cases}.$$

(2)

Using the parameters of the bond nature $m$, the atomic coordination $z$, the bond length $d$ and the bond energy $E$, we can predict the bond energy ratio $\gamma$, the local bond strain $\varepsilon_z$, the bond energy density $\delta E_d$ and the atomic cohesive energy $\delta E_c$:

$$\begin{cases} \gamma = E_i/E_b = C_i^{-m} = \left(\Delta E_v(12) + \Delta E_v'(i)\right)/\Delta E_v(12) & \text{(bond energy ratio)} \\ \delta E_c(z) = z_i E_i / z_b E_b = z_{ib} C_i^{-m} - 1 = z_{ib}\gamma - 1 & \text{(atomic cohesive energy)} \\ \delta E_d(z) = (E_i/d_i^3)/(E_b/d_b^3) - 1 = C^{-m-3} - 1 = \gamma^4 - 1 & \text{(bonding energy density)} \\ \varepsilon_z = d_i/d_b - 1 = \gamma^{-1} - 1 & \text{(local bond strain)} \end{cases}$$

(3)

$z_{ib} = z_i/z_b$ is an atomic coordination ratio with $z_b = 12$ being the bulk standard. The bond energy ratio $\gamma$ is represents the bond energy strength. If the bond energy ratio $\gamma > 1$, the bond energy is become stronger and electronic quantum entrapment; otherwise, $\gamma < 1$, the bond energy is become weaker and electronic polarization. The local bond strain $\varepsilon_z$ is represents contract ratio of the bond length. The bond energy density $\delta E_d$ is represents strong and weak localization of electrons. The atomic cohesive energy $\delta E_c$ is atomic binding strength, and the product of atomic coordination $z$ and bond energy $E_i$.

## 3. Results and discussion

**3.1 Geometric structures and energy shifts of Y 4$p$ level**

Before studying the adsorption Y atoms on Li(110) surfaces, we optimize geometric structures. We found the 2D structures of Y metal formation over Li metal substrates, as is shown in the **Fig. 1**. We found the geometric structures of Triangle that the maximum height ($h_i$) between the adsorption Y atoms and Li(110) surface is 2.73 Å, and initial height is 2.48 Å, as shown in the Table 1. The result indicates that the height of triangles structures is increase 12.76%. According to the BOLS notion, the height is larger, and the layer and layer interacted is weaker. As for geometric structures of nonagon, the minimum height is 1.66 Å. which is agreement with BOLS notion of the larger energies shifts with the shorter distance, as shown in the **Table 1**.

In the XPS experimental measure, the atomic CN is using for determinations between surface and bulk atoms[18]. Thus, atomic CN is importance parameter of geometric structures. In Section 2.2, we have the relation of energy shifts $\Delta E_v(i)$ and the atomic CN $z$. Using Eq. (3), we can calculated the atomic CN $z$. **Fig.** 2 shows DFT calculations of the 4$p$-orbit DOS for structures hexagon, nonagon, solid hexagonal, quadrangle and triangle. The results shown that the peaks of the energy shift are toward higher binding energy as the atomic CN $z$ was reduced.

DFT calculations and BOLS notion were not only enables reproduction of the energies shifts and atomic CN $z$ but also quantitative information on the bonding identities of the bond energy ratio $\gamma$, the local bond strain $\varepsilon$, the bond energy density $\delta E_d$ and the atomic cohesive energy $\delta E_c$. A more detailed understanding of the atomic bonding identities can be obtained from energies shifts of 4$p$ energy level. Using Eq. (3), we can predict bonding identities of $\gamma$, $\varepsilon_i$, $\delta E_c$ and $\delta E_d$ for structures hexagon, nonagon, solid hexagonal, quadrangle and triangle of adsorption Y atoms on Li(110) surfaces, as is shown in the Table 1.

**3.2 Magnetic properties and Electronic Structures of adsorption Y atoms on Li(110) surfaces**

On the basis of the stable structures of hexagon, nonagon, solid hexagonal, quadrangle and triangle, we systematically investigate their magnetic and electronic properties. To understand the magnetism of adsorption Y atoms on Li(110) surfaces, partial DOS are shown in the Fig. 3 and 4. The difference between the occupied spin-up and spin-down band is the magnetic moment; the magnetic moment is given in the Table 1. The sequence of magnetic moment of adsorption Y atoms on Li(110) surfaces is hexagon > nonagon > solid hexagonal > quadrangle > triangle, respectively.

The Fig. 3 and 4 showed the spin-resolved partial DOS and difference between spin-up and spin-down states of adsorption Y atoms on Li(110) surfaces with hexagon, nonagon, solid hexagonal, quadrangle and triangle structures. Magnetic behaviour of adsorption Y atoms on Li(110) surfaces, we found that the Li 1$s$-orbit of the spin-up and spin-down states is equal, as is shown in the Fig 3. The partial DOS is shown the Y 4*s* level that the two spin channels are unequally distributed which means the magnetic behaviour is from the adsorption Y atoms. Their shows that the component of Li metal is not have magnetic behaviour in these systems; the magnetic behaviour is from the Y metal.

The Fig. 4 is shown the valence band of the spin-resolved partial DOS for the *s*-, *p*-, and *d*- orbitals. It's shown that *s*- orbitals of spin-up and spin-down states are equal distribution, *p*-, and *d*- orbitals of spin-up and spin-down are unequally distribution. We can find that origin of the two spin channels are unequally of magnetic moment which are from 4 orbitals, it's the valence band of *p*-, and *d*- orbitals and core band of 4*s*- and 4*p*- orbitals of Y atoms.

We analyse the magnetic behaviour of these systems by calculating core band of the spin shifts of 4*p* level $\Delta E_{4p}(s)= E_{spin}(up)- E_{spin}(down)$. Furthermore, Fig. 5 showed 4*p*-orbit of adsorption Y atoms have spin shifts for hexagon, nonagon, solid hexagonal, quadrangle and triangle structures are 0.01 eV, 0.05 eV, 0.16 eV, 0.19 eV, 0.33eV, respectively. This result is consistent with the magnetic moment 0.28μB, 1.04μB, 2.81μB, 5.54μB, 6.66μB, respectively. The larger spin orbit splitting has the larger spin shifts.

### 3.3 Mulliken population analysis and deformation charge densities

A more detailed understanding of the electrons and bonds dynamic process can be obtained from the charge transfer of Li and Y atoms. We using the Mulliken population analysis[19] estimate the charge transfer between Li and Y atoms, as shown in Table 1. We found the Y atom gets electrons from the Li atoms. The result is shown that the adsorption Y atoms on Li(110) surfaces induced the Y 4*s*- and 4*p*-orbit level positive energy shifts and Y-Y metal bond energy is become stronger ($\gamma >1$). We found that the formation of 2D structures of the Y metal on the Li(110) surface is mainly due to changes in the electron distribution and Y-Y atomic bonding energy enhance. In Fig. 6 are plotted the deformation charge densities[20] of 2D structures of the Y metal on the Li(110) surface; the blue charge density is negative, meaning that electrons are lost from those parts, and the red charge density is positive, meaning that electrons are gathered into those parts of the structures. As the reference of triangle structure, strong and weak localization of electrons is consistent with change of bond energy density of hexagon > solid hexagonal > nonagon > quadrangle, as shown in the Table 1 and the Fig.6.

4. **Conclusion**

In summary, we have performed BOLS notation and DFT calculations on the magnetic and electronic properties of adsorption Y atoms on Li(110) surfaces. The consistency between BOLS notation and DFT calculations clarified that:

(i) We find that the adsorption Y atoms on Li(110) surfaces form 2D geometric structures of hexagon, nonagon, solid hexagonal, quadrangle, triangle structure. Energy shifts of Y 4$p$ level due to the under-coordination atoms enhance the bond energy. Furthermore, atomic bond energy may be attributed to the local bond strain, atomic cohesive energy and bond energy density.

(ii) DFT calculated total magnetic moment ***M*** for geometric structures hexagon, nonagon, solid hexagonal, quadrangle, triangle are 0.28$\mu_B$, 1.04$\mu_B$, 2.81$\mu_B$, 5.54$\mu_B$, 6.66$\mu_B$, respectively. In accordance with magnetic moment, the energy difference of 4$p$ level spin $\Delta E_{4p}(s)$ are 0.01 eV, 0.05 eV, 0.16 eV, 0.19 eV and 0.33 eV, respectively. We find that core band is equally important of the spin state.

These findings may provide a new avenue for 2D metal nanomaterials in-depth theoretical and experimental investigations.

**Acknowledgment**

Financial support was provided by the NSF (Grant No. 11747005), the Science and Technology Research Program of Chongqing Municipal Education Commission (Grant No. KJ1712299), and Yangtze Normal University (Grant No. 2016XJQN28 and 2016KYQD11).

**Table and figure captions:**

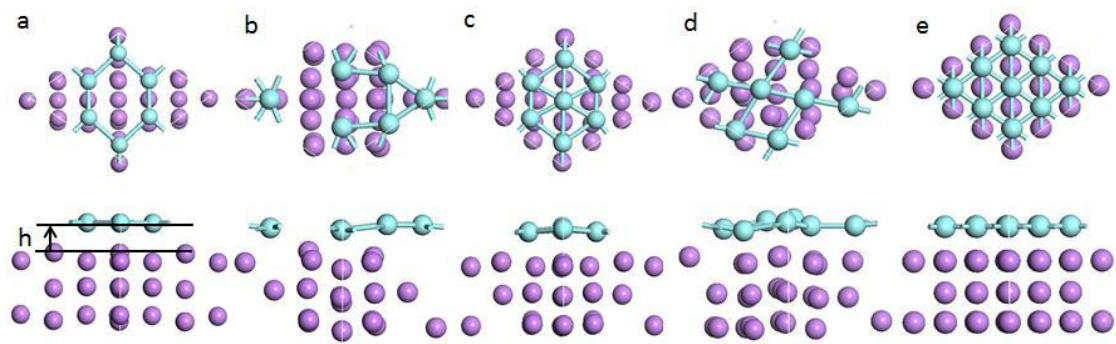

**Fig. 1** Geometric structures (a) hexagon, (b) nonagon, (c) solid hexagonal, (d) quadrangle (e) triangle of adsorption Y atoms on Li(110) surfaces. Blue and Violet correspond to Y and Li atoms, respectively.

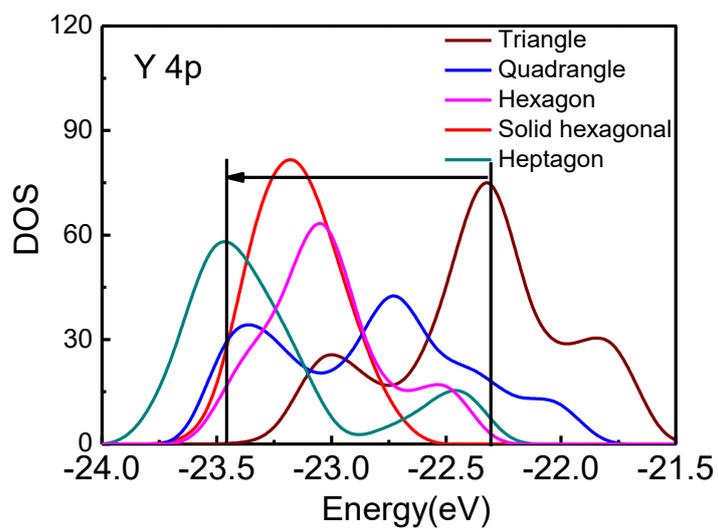

**Fig. 2** Energy shifts of Y 4$p$ level.

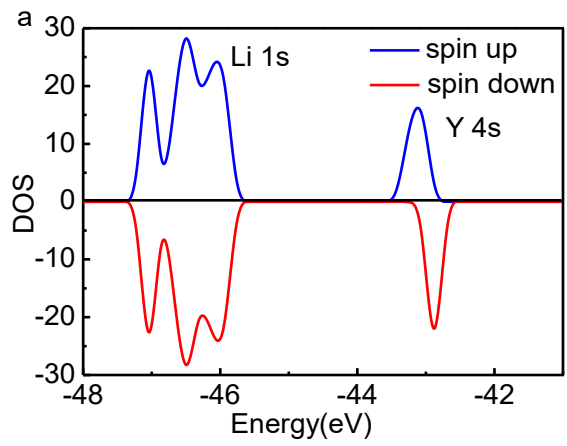

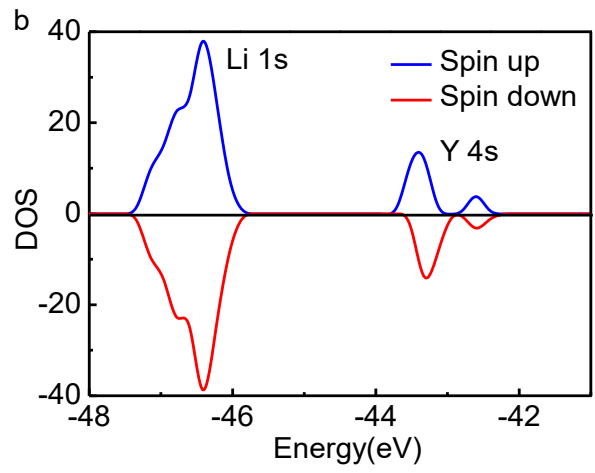

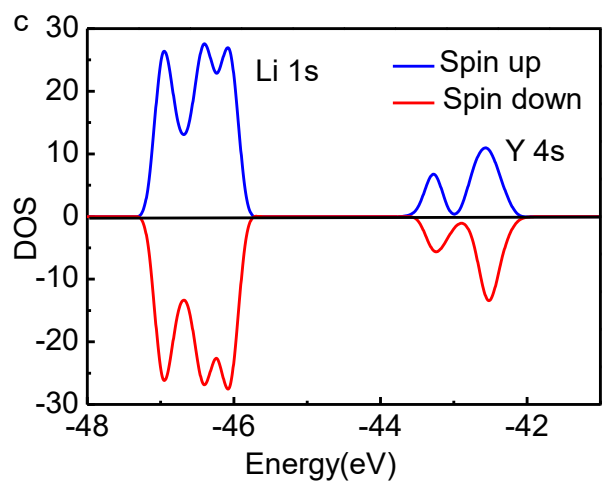

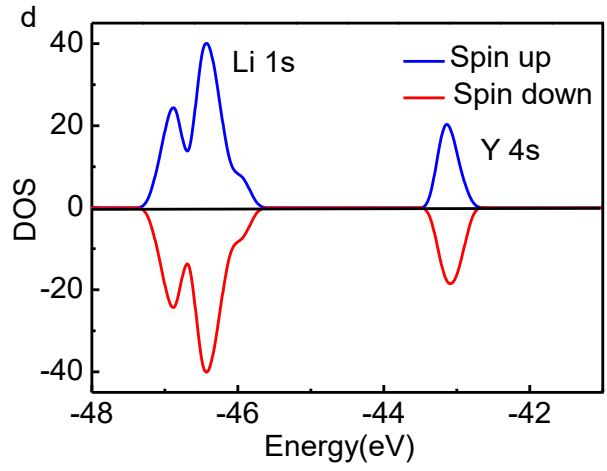

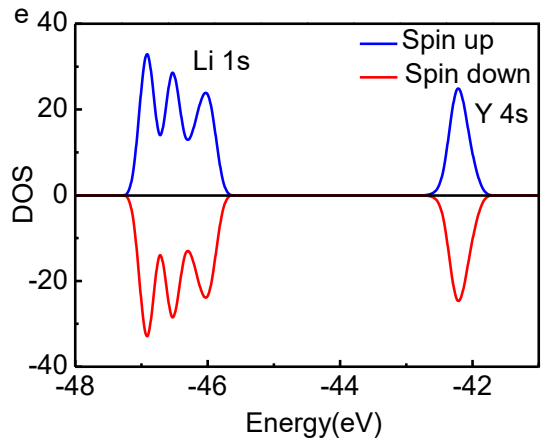

Fig. 3 Spin up and spin down for(a) hexagon, (b) nonagon, (c) solid hexagonal ,(d) quadrangle (e) triangle structures of Y 4*s* level and Li 1*s* level.

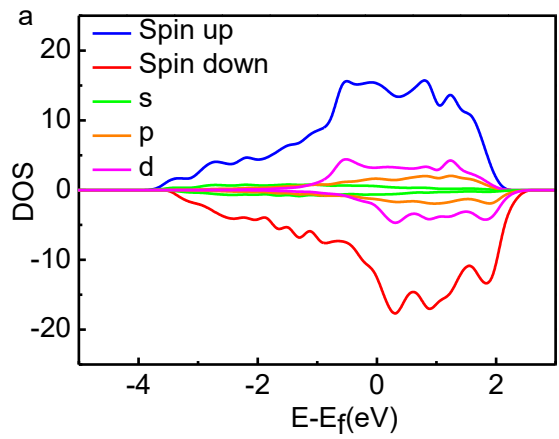

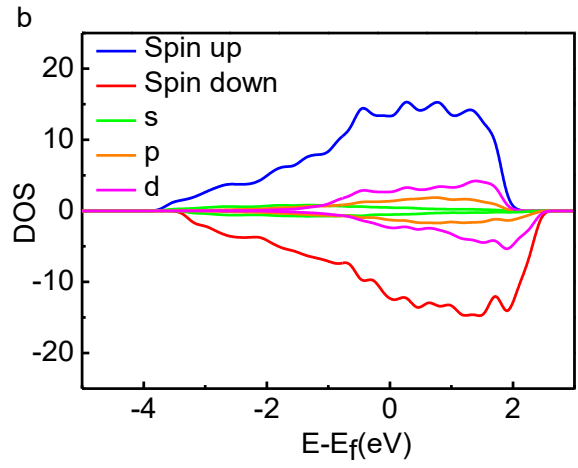

b

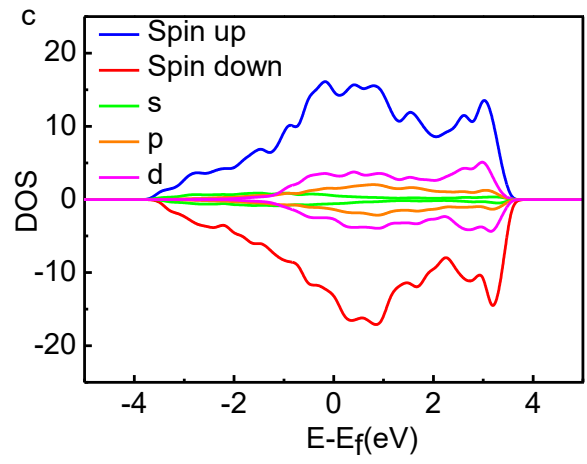

c

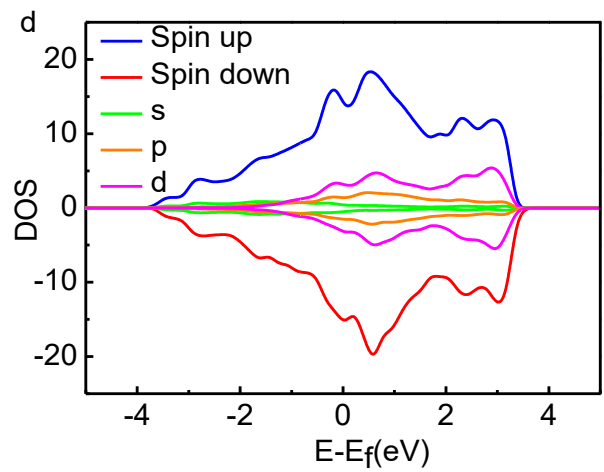

d

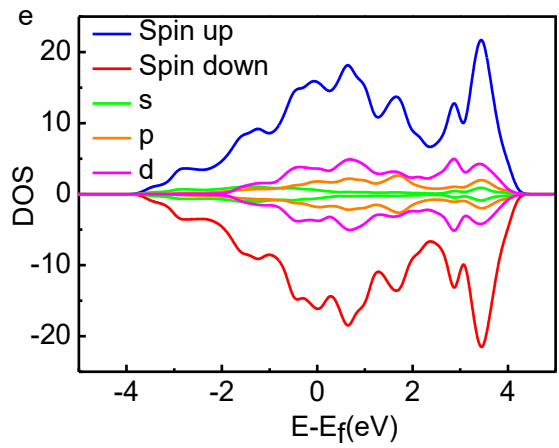

Fig. 4 PDOS for (a) hexagon, (b) nonagon, (c) solid hexagonal ,(d) quadrangle (e) triangle structures of Y metal.

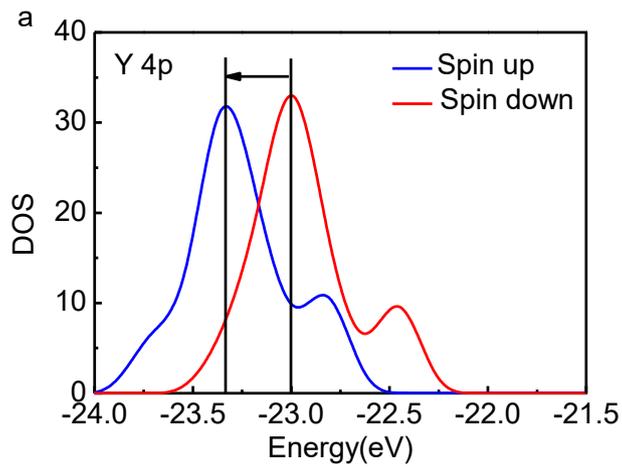

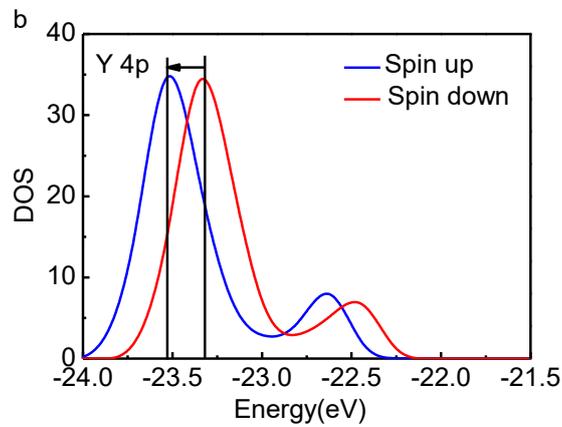

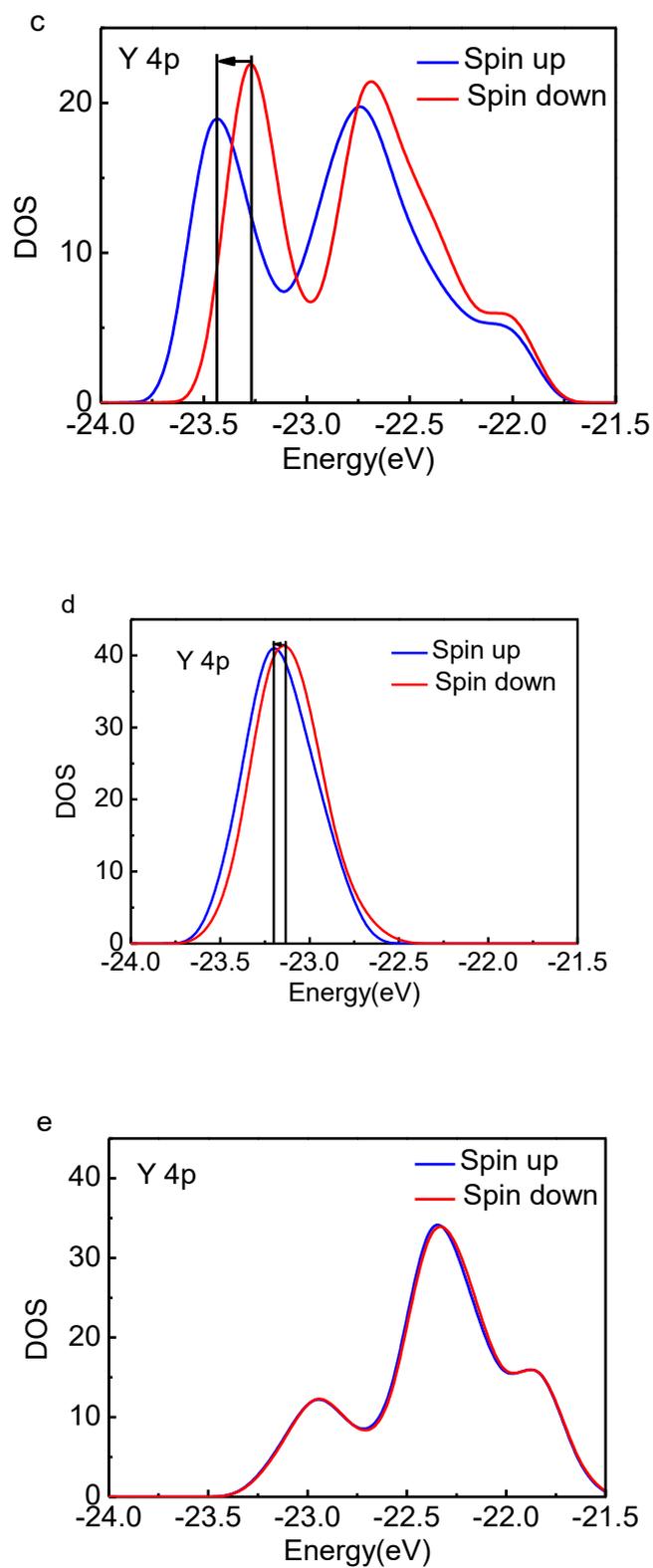

**Fig. 5** Spin up and spin down for (a) hexagon, (b) nonagon, (c) solid hexagonal, (d) quadrangle (e) triangle structures of Y 4*p* level.

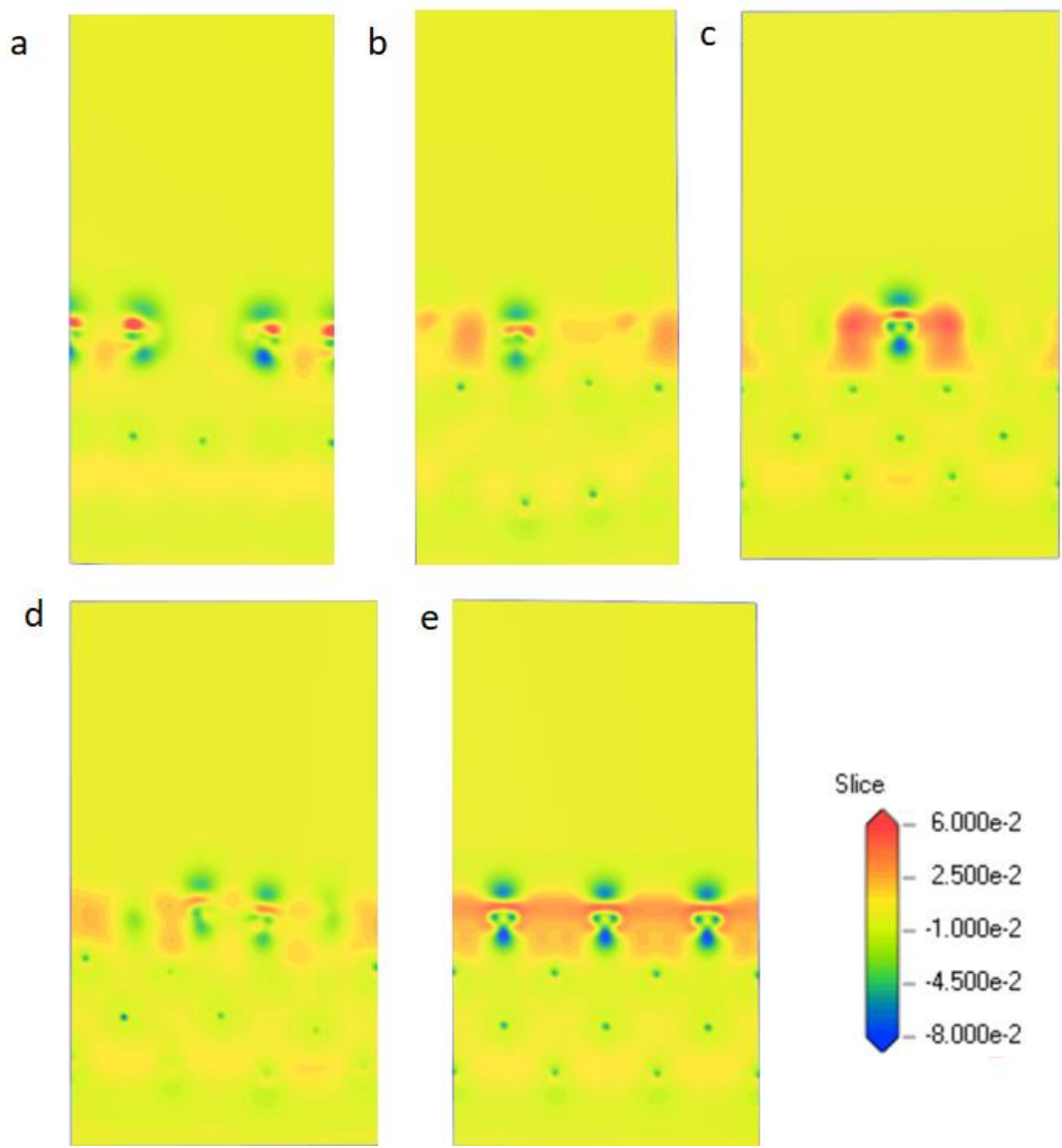

**Fig. 6** Deformation charge densities (a) hexagon, (b) nonagon, (c) solid hexagonal ,(d) quadrangle (e) triangle structures

**Table 1** The energy difference of spin $\Delta E_{4p}(s) = E_{spin}(up) - E_{spin}(down)$ (eV), binding energy of 4p level $E_{4p}(i)$, energy shifts $\Delta E_{4p}(i) = E_{4p}(i) - E_{4p}(12)$ (eV), the average effective CN $z$, the bond energy ratio $\gamma(\%)$, total magnetic moment M($\mu_B$), height $h_i$(Å) represents the shortest distance between the adsorption Y atoms and Li(110) surface, local bond strain $-\varepsilon_i(\%)$, atomic cohesive energy $-dE_c(\%)$ and bond energy density $\delta E_d(\%)$.

| Geometric structures | $E_{4p}(i)$ | $\Delta E_{4p}(z)$ | $z$ | M | $h_i$ | $\Delta E_{4p}(s)$ | $\gamma$ | $-\varepsilon_z(\%)$ | $\delta E_d(\%)$ | $-\delta E_c(\%)$ | Charge[a] (Y) |
|---|---|---|---|---|---|---|---|---|---|---|---|
| Triangle | 22.33 | 0 | 12 | 0.28 | 2.73 | 0.01 | 1 | 0 | 0 | 0 | -2.56 |
| quadrangle | 22.72 | 0.39 | 4.71 | 1.04 | 2.26 | 0.05 | 1.11 | 9.63 | 49.92 | 56.52 | -2.46 |
| Solid hexagonal | 23.19 | 0.86 | 2.94 | 2.81 | 2.47 | 0.16 | 1.23 | 19.02 | 132.56 | 69.74 | -2.39 |
| nonagon | 23.47 | 1.14 | 2.46 | 5.54 | 1.66 | 0.19 | 1.31 | 23.75 | 195.75 | 73.09 | -2.50 |
| Hexagon | 23.04 | 0.71 | 3.31 | 6.66 | 2.19 | 0.33 | 1.19 | 16.24 | 103.20 | 67.04 | -2.62 |

[a]Negative sign indicates charge gain, otherwise, a charge loss occurs.

**References:**

1. W. N. Kang, H.-J. Kim, E.-M. Choi, C. U. Jung and S.-I. Lee, Science **292** (5521), 1521-1523 (2001).

2. H. Duan, N. Yan, R. Yu, C.-R. Chang, G. Zhou, H.-S. Hu, H. Rong, Z. Niu, J. Mao, H. Asakura, T. Tanaka, P. J. Dyson, J. Li and Y. Li, Nature Communications **5**, 3093 (2014).

3. X. Huang, S. Tang, X. Mu, Y. Dai, G. Chen, Z. Zhou, F. Ruan, Z. Yang and N. Zheng, Nature Nanotechnology **6**, 28 (2010).

4. C. Gao, Z. Lu, Y. Liu, Q. Zhang, M. Chi, Q. Cheng and Y. Yin, Angewandte Chemie International Edition **51** (23), 5629-5633 (2012).

5. S. R. Beeram and F. P. Zamborini, Journal of the American Chemical Society **131** (33), 11689-11691 (2009).

6. S. Woo, K. Litzius, B. Krüger, M.-Y. Im, L. Caretta, K. Richter, M. Mann, A. Krone, R. M. Reeve, M. Weigand, P. Agrawal, I. Lemesh, M.-A. Mawass, P. Fischer, M. Kläui and G. S. D. Beach, Nature Materials **15**, 501 (2016).

7. B. Heinrich and J. F. Cochran, Adv. Phys. **42** (5), 523-639 (1993).

8. T. Ling, J.-J. Wang, H. Zhang, S.-T. Song, Y.-Z. Zhou, J. Zhao and X.-W. Du, Adv. Mater. **27** (36), 5396-5402 (2015).

9. Y. Chen, Z. Fan, Z. Zhang, W. Niu, C. Li, N. Yang, B. Chen and H. Zhang, Chem. Rev. **118** (13), 6409-6455 (2018).

10. J. Zhao, Q. Deng, A. Bachmatiuk, G. Sandeep, A. Popov, J. Eckert and M. H. Rümmeli, Science **343** (6176), 1228-1232 (2014).

11. S. Zhang, Z. Yan, Y. Li, Z. Chen and H. Zeng, Angewandte Chemie International Edition **54** (10), 3112-3115 (2015).

12. K. R. E., *The history and use of our earth's chemical elements: a reference guide*. (Greenwood Publishing Group, 2006).

13. C. Q. Sun, *Relaxation of the Chemical Bond*. (Heidelberg New York Dordrecht London Singapore 2014 ).

14. John P. Perdew, Kieron Burke and M. Ernzerhof, Phys. Rev. Lett. **77** (18), 3865-3868 (1996).

15. M. Bo, Y. Guo, Y. Huang, Y. Liu, Y. Wang, C. Li and C. Q. Sun, RSC Adv. **5** (44), 35274-35281 (2015).

16. Y. Wang, Y. Nie, L. Wang and C. Q. Sun, J. Phys. Chem. C **114** (2), 1226-1230 (2010).

17. M. Bo, Y. Guo, Y. Liu, Y. Wang, C. Q. Sun and Y. Huang, RSC Advances **6** (10), 8511-8516 (2016).

18. M. Bo, Y. Wang, Y. Huang, X. Yang, Y. Yang, C. Li and C. Q. Sun, Applied Surface Science **320**, 509-513 (2014).

19. R. S. Mulliken, The Journal of Chemical Physics **23** (10), 1841-1846 (1955).

20. C. S. a. K. Wang, B. M., Phys. Rev. B **24** (6), 3393-3416 (1981).